\documentclass[prd,nofootinbib]{revtex4}
\usepackage{amsmath, amssymb, graphicx, comment, amsthm}

\newcommand{\none}{\nonumber \\}
\newcommand{\half}{\frac{1}{2}}
\newcommand{\tOmega}{\tilde{\Omega}}

\def\be {\begin{equation}}
\def\ee  {\end{equation}}
\def\bea {\begin{eqnarray}}
\def\eea {\end{eqnarray}}

\begin{document}
\title{Two Dimensional Gravity as a modified Yang-Mills Theory}

\author{Jack Gegenberg}

\affiliation{Department of Mathematics \& Statistics, University of
New Brunswick \\ Fredericton, New Brunswick, E3B 5A3, Canada}

\author{Gabor Kunstatter}

\affiliation{ Department of Physics, University of Winnipeg and Winnipeg Institute for Theoretical Physics, Winnipeg, Manitoba,Canada R3B 2E9}

\begin{abstract} %
We study a  deSitter/Anti-deSitter/Poincare Yang-Mills
theory of gravity in d-space-time dimensions in an attempt to retain the best
features of both general relativity and Yang-Mills theory: quadratic
curvature, dimensionless coupling  and background independence. We derive the
equations of motion for Lie algebra valued scalars and show that in the
geometric optics limit they traverse geodesics with respect to the Lorentzian
geometry determined by the frame fields. Mixing between components appears to
next to leading order in the WKB approximation. We then restrict to two
space-time dimensions for simplicity,  in which case the theory reduces to the well known Katanaev-Volovich model. We complete the Hamiltonian analysis of
the vacuum theory and use it to prove a generalized Birkhoff theorem. There
are two classes of solutions: with torsion and without torsion. The former
are parametrized by two constants of motion, have event horizons for certain
ranges of the parameters and a curvature singularity. The latter yield a
unique solution, up to diffeomorphisms, that describes a space  constant
curvature . \end{abstract} \maketitle \tableofcontents \section{Introduction}

General Relativity is often called a gauge theory of the gravitational field,
but it is not a gauge theory of the Yang-Mills type.  In the latter, the
action functional $S_{YM}[A]$ is quadratic in the curvature of a connection
$A$ of principle bundle over the spacetime manifold $(M_D,{\bf g})$, where
${\bf g}$ is a {\it given} non-dynamical Lorentzian metric on $M_D$:
\be
S_{YM}[A]=\frac{1}{8 g_{YM}^2}\int_{M_D} d^D x
\sqrt{-g}g^{\mu\nu}g^{\alpha\beta}F^A_{\mu\alpha} F^B_{\nu\beta} h_{AB}.
\label{eq:action1}
\ee
In the above, $g_{YM}$ is the gauge coupling constant.  It has dimension Length${}^{\frac{D}{2}-2}$, and hence
is dimensionless in 4D; the $g^{\mu\nu}$ are the contravariant
components of the metric tensor ${\bf g}$; the $F^A_{\mu\alpha}$ are the
components of the curvature of the connection; the indices $A,B=1,2,...,n$
are in the adjoint representation of the gauge group; and finally $h_{AB}$
are components of the Cartan-Killing metric of the group. It is the fact that
$g_{YM}^2$ is dimensionless in 4-D which permits Yang-Mills gauge theories to be
perturbatively renormalizable

By contrast, the Einstein-Hilbert action of General Relativity is linear in
the curvature of the Christoffel connection of the Lorentzian metric.  There
is no background:  the metric ${\bf g}$ is dynamical. Moreover,  the coupling
constant in Einstein gravity has dimension Length${}^{-2}$ in four spacetime
dimensions.  It is this which has stalled progress in constructing quantum
gravity starting from Einstein's theory.

 It was perhaps Townsend \cite{townsend} who first highlighted the fact that
 the gravitational constant has the dubious distinction of being the only
 dimensionful fundamental constant (the others being $\hbar$ and $c$) that is
 tied to a specific dynamical theory. He therefore suggested that  the gravitational constant $G$ should
 somehow be linked directly to the structure of spacetime. This could be
 achieved by replacing the Poincare group as a potential local gauge symmetry
 of gravity by  the deSitter group, which necessarily entails a dimensionful
 constant. With this as motivation, he proceeded to consider a Yang-Mills
 type Lagrangian for gravity with the deSitter group as gauge group.

Besides those of Townsend, there have in fact been many attempts to construct
a Yang-Mills type gravitational theory.  The first was by Weyl almost one
hundred years ago, and the goal was to unify gravity with electromagnetism
\cite{weyl}.   Work in the 70's and 80's, inspired by the work of Utiyama,
Yang and Mills on non-Abelian gauge theories, constructed Yang-Mills type
theories with gauge groups associated with gravity, for example the Poincare,
DeSitter/anti DeSitter and Conformal groups \cite{early}. More recently, J.
T. Wheeler\cite{wheeler} and collaborators have worked on 4D Yang-Mill
gravity, with the conformal group SO(4,2) as the gauge group, while H.-Y.
Guo\cite{guo} and his collaborators have tackled the de Sitter case.

In the first order formalism of Einstein gravity- the so-called
Einstein-Cartan action- the equations of motion force the torsion to be zero.
In Yang-Mills gravity this does not happen.  Generically, the spacetime
geometry has non-vanishing torsion as well as (quasi-)Riemannian curvature
\cite{early,guo}.  The consequences of this for the viability of such
theories is still an open question.  We note here the result of
\cite{ymflow} that torsion de-stabilizes anti-de Sitter 2+1 dimensional
spacetime.

In spite of its quantum motivation, little progress has been made in
quantizing Yang-Mills gravity. In fact, to date, there has been no canonical
analysis of such theories, a necessary first step towards understanding the
quantum theory. In this paper we begin to close this gap.   After a general
discussion that includes a discussion of the coupling to matter showing
that, to leading order in the geometric optics limit, Higgs fields propagate
along geodesics of the Lorentzian geometry, we will undertake the
construction of the canonical form of a toy model of Yang-Mills gravity,
wherein the spacetime is two dimensional, and the gauge group is the lineland
version of de Sitter/anti-de Sitter/Poincare gravity, that is,
SO(2,1)/SO(1,2)/ISO(1,1). In this case, the Lagrangian reduces to a special case of the Katanaev-Volkov model\cite{kv86}, which was extensively studied in a somewhat different context in the 1990's\cite{kv90s}.
 We will solve the Hamiltonian equations of motion
for the vacuum case, finding  in the case of zero torsion that the solutions
are equivalent to those of Jackiw-Teitelboim dilaton gravity \cite{jt}.

\section{General Formalism}
\subsection{Algebra and Action}
In this section we outline the general procedure for constructing a gauge theory of gravity in a $D$-dimensional spacetime.
We note here the record of such attempts sampled in \cite{townsend, weyl, early, wheeler,guo}.  The `kinematical' gauge group
associated with such a theory is one SO(D,1)/SO(D-1,2)/ISO(D-1,1), corresponding to positive/negative/zero cosmological constant.
The generators $J_A=(J_a,F_{ab})$ with $ A=0,1,...,D;{}{} a,b=0,1,...D-1$ obey
\be
[J_a,J_b]=-2\eta_{DD}J_{ab};
\ee
\be
[J_a, J_{bc}]=-\half\left(\eta_{ac}J_b-\eta_{ab}J_c\right);
\ee
\be
[J_{ab},J_{cd}]=-\half\left(\eta_{ac}J_{bd}+\eta_{bd}J_{ac}-\eta_{bc}J_{ad}-\eta_{ad}J_{bc}\right),
\ee
where $\eta_{ab}$ is the $(D)$-dimensional Minkowski metric.  If $\eta_{DD}=1,-1,0$, then the gauge group is, respectively, SO(D,1)/SO(D-1,2)/ISO(D-1,1).

The gauge potential is decomposed according to
\be
A_\mu=\lambda e^a_\mu J_a +\omega_\mu^{ab} J_{ab},
\ee
where the constant $\lambda$ has dimension $L^{-1}$, so that with the vielbein $e^a_\mu$ dimensionless,
the spin-connection $\omega^{ab}_\mu$ and the gauge potential $A_\mu$ have dimension $L^{-1}$. The generator of
translations, i.e. the 2-momentum, $P_a$ has dimension $L^{-1}$ and is related to the above via:
\be
P_a = \lambda J_a
\ee
When working in terms of $P_a$, $\lambda$ appears directly in the commutator algebra as opposed to the
definition of the gauge potential. It is for this reason  that Townsend\cite{townsend} considered it to
be a property of the spacetime structure, rather than a coupling constant.\\
Note that although it would be natural to identify $\lambda$ with the dimensionally appropriate power of the $g_{YM}$, up to a dimensionless number of order unity, in order to keep things as general as possible we keep them distinct in what follows.

We note here that the structure constants for any of the DS/ADS/Poincare gauge groups have structure constants
\bea
f^{[cd]}{}_{ab}&=&-2\eta_{DD}\delta^{[c}_a\delta^{d]}_b;\none
f^d{}_{a[bc]}&=&-\delta^d_{[b}\eta_{c]a};\none
f^{ef}{}_{[ab][cd]}&=&-2\delta^{[e}_{[a}\eta_{b][d}\delta^{f]}_{c]}.
\eea
The Cartan-Killing metrics on the gauge groups, defined by $h_{ij}:=2 f^k{}_{il} f^l{}_{jk}$ are all of the form
\be
h_{ab}=-2D\eta_{DD}\eta_{ab}; h_{[ab][cd]}=-D\left(\eta_{ac}\eta_{bd}-\eta_{bc}\eta_{ad}\right).
\ee

The field strength is defined as usual:
\bea
F=\half F^{ij}_{\mu\nu}dx^\mu\wedge dx^\nu :&=&dA+\half[A,A]\none
&=& \lambda T^a J_a+\Omega^{ab} J_{ab}.
\eea
Thus $[F]=L^{-2}$.
The Lie algebra valued 2-forms $\Omega$ and $T$ are respectively the `curvature plus volume element' and torsion of the spin connection $\omega$:
\bea
\Omega^{ab}&:=& d\omega^{ab}+\omega^{ac}\wedge \omega_c{}^b-\lambda^2\eta_{DD}e^a\wedge e^b;\label{curv}\\
T^a&:=&de^a +\omega^a{}_b\wedge e^b.\label{torsion}
\eea
In the above, indices $a,b,...=0,1,2,D-2$ are raised and lowered by the Minkowski metric $\eta_{ab}$.
and e.g. $F^a:=\frac{1}{2} F^a_{ij} dx^i\wedge dx^j$.  Thus $\left[T^a_{\mu\nu}\right]=L^{-1}$ and $\left[\Omega^{ab}_{\mu\nu}\right]=L^{-2}$.
Most importantly, the `background metric',
\be
g_{\mu\nu}:= \eta_{ab} e^a_\mu e^b_\nu,
\ee
is not fixed, but is subject to the dynamics determined by the equations of motion for the gauge field.

The action can be written explicitly in the form $S=S_{EH}+S_1$, where
\bea
S_{EH}:&=&\frac{D\lambda^2}{2 g_{YM}^2}\int d^D x\sqrt{-g}\left(R-\lambda^2\eta_{
DD}\frac{D(D-1)}{2}\right);\\
S_1:&=&-\frac{D}{4 g_{YM}^2}\int d^D x\sqrt{-g}\left(\frac{K}{2}+ \lambda^2\eta_{DD}T_{a\mu\nu}T^{a\mu\nu}\right),
\eea
where $K:=R^{ab\mu\nu}R_{ab\mu\nu}$. \\

 Comparing the term $S_{EH}$ to the usual Einstein-Hilbert action, we find that the Newton gravitational constant in $D$ dimensions, $G_D$, (in units where the speed of light is one)  is related to the gauge coupling constant $G$ and the scale factor $\lambda$ by
\be
G_D=\frac{ g_{YM}^2}{8\pi D \lambda^2}.
\ee
In general $G_D$ has dimensions of $L^{D-2}$, so that it is dimensionless in 2 spacetime dimensions. We also remark again that $D=4$ is also special in that $g_{YM}$ is dimensionless.

To close this section, we consider the issue of the background metric ${\bf g}$.  In the following,
as in most of the literature on Yang-Mill gravity, the `background' $g_{\mu\nu}$ will not really be a
background, but is rather, dynamical, via $g_{\mu\nu}:=\eta_{ab} e^a_\mu e^b_\nu$.  One pays a price
for this, however, in that the gauge transformations generated by the `translations' $J_a$ no longer preserve the action:  some of the gauge symmetry is broken.

\subsection{Adding Matter}

Torsion theories are distinguished by the fact that the field content describes more than one kind of geometry.
There is curvature associated with the Riemannian metric used to raise and lower indices, and there is also the
Riemann-Cartan connection and associated curvature. When the torsion is non-zero, the metric compatible with the
Riemann-Cartan connection is not the same as the Riemannian metric constructed out of the fierbeins/vielbeins.
The only way to decide which geometry is relevant in a particular physical context is to look at matter couplings.

It is straightforward to add most forms of matter using the principle of minimal couple. Only spinors will
couple directly to the torsion, whereas all other matter Lagrangians will just depend on $e^a{}_\mu$.
Here we consider a Higgs-like scalar $\phi^A(x)$ that takes its values in the adjoint representation and
couples to the vacuum action via $S=S_{YM}+S_{higgs}$, where
\be
S_{higgs}=\int_{M_2} d^2x\sqrt{-g} g^{\mu\nu}h_{ij} D_\mu\phi^i D_\nu \phi^j.
\ee.

Thus the matter field obeys the gauge covariant wave equation
\be
D_\mu D^\mu \phi^i=0.
\ee


In the geometric optics limit, a wave field has approximately constant amplitude, but varying phase.
Thus for a Higgs type of matter we write
\be
\phi^j=R^j e^{iS^j/\hbar}
\ee
Note that the Lie algebra index $j=0,1,2$ is not summed over here, or subsequently.

Now the geometric optics limits has particles traveling with momenta $k^{(i)}_\mu=\partial_\mu S^i$ orthogonal
to the constant surfaces $S^i(t,x)=const$.   Also, $\nabla_\mu$ is the Lorentzian covariant derivative with
respect to the background metric $g_{\mu\nu}=\eta_{ab}e^a_\mu e^b_\nu$.    We assume that the amplitudes $R^j$
are  slowly varying compared to the phase.  The wave equation $D^\mu D_\mu \phi^j=0$ becomes, after dropping
terms in $\partial_\mu R^j$ and keeping only terms of leading and subleading orders ($1/\hbar^2, 1/\hbar$, respectively)
\be
0=-\frac{1}{\hbar^2}e^{iS^j/\hbar} R^j k^{(j)\mu} k^{(j)}_\mu+\frac{i}{\hbar}\left(e^{i S^j/\hbar}R^j\nabla^\mu k^{(j)}_\mu +
2 e^{iS^l/\hbar}f^j{}_{kl}A^{(k)\mu} k^{(l)}_\mu R^l\right).
\ee
Hence, to leading order
\be
C^j:=-R^j k^{(j)\mu} k^{(j)}_\mu=0 \label{eq:ray}
\ee
Note that this expression is real.  The gauge covariant derivative of $C^j$ reduces to the
partial derivative.  Thus, to leading order $k^{j\nu}\nabla_\mu k^j_\nu=0$.  Using the
smoothness of the phase $S^j$ in order to change the order of partial differentiation
we find that $k^{j\nu}\nabla_\nu k^j_\mu=0$.  Thus to leading order the trajectories
are null geodesics of the Lorentzian geometry compatible with the frame-field $e^i_\mu$ on spacetime.

The subleading terms are pure imaginary terms, and hence
\be
e^{i S^j/\hbar}R^j\nabla^\mu k^{(j)}_\mu +2 e^{iS^l/\hbar}f^j{}_{kl}A^{(k)\mu} k^{(l)}_\mu R^l=0
\ee
The latter are more complicated because of the relative phase factor.

\section{1+1 Dimensions: Action and Covariant Equations of Motion}

Things simplify quite a bit in 1+1 dimensions.
The group is $SO(2,1),SO(1,2),ISO(1,1)$ respectively, for $k:=-\eta_{22}=-1,+1,0$ with generators $J_a,J$, and algebra:
\bea
[J,J] &=& 0\\
\left[J_a,J_b\right] &=& k \epsilon_{ab}J  \nonumber\\
\left[J,J_a\right]&=&\epsilon_a{}^b J_b
\eea
Note that $J$ generates an Abelian one dimensional subalgebra.
\be
A = \lambda e^a J_a + \omega J
\ee

As before we split the curvature into
\be
F= \Omega J + \lambda T^a J_a
\ee
where $\Omega:=d\omega+\frac{k}{2}\lambda^2\epsilon_{ab}e^a\wedge e^b$ and $T^a:=de^a-\epsilon^a{}_b\omega\wedge e^b$. That is:
\bea
\Omega_{\mu\nu}&=& \partial_\mu\omega_\nu - \partial_\nu\omega_\mu + k \lambda^2 V_{\mu\nu}\\
T^a_{\mu\nu}&=&\partial_\mu e^a_\nu-\partial_\nu e^a_\mu -\epsilon^a{}_b(\omega_\mu e^b_\nu-\omega_\nu e^b_\mu)
\eea
with $V_{\mu\nu}:= \epsilon_{ab} e^a{}_\mu e^b{}_\nu$. To recover the expressions from the
previous section, in an arbitrary number of dimensions, we replace $\omega=\half\epsilon^{ab}\omega_{ab}$.  Note that $V^{\mu\nu}V_{\mu\nu} = -2$ and  $k:=-\eta_{22}$.
The cartan metric $h_{ij}=h_{ji}$ is defined to be:
\be
h_{ij} := -2 f_{ki}{}^l f_{lj}{}^k
\ee
with components:
\bea
h_{22} &=& -4\\
h_{ab} &=& -4k\eta_{ab}\\
h_{a2} &=& 0
\eea
The action  Eq.(\ref{eq:action1}) becomes:
\be
S_{YM} = \frac{1}{4\lambda^2} \int d^2 x \sqrt{-g} \left(-\tilde{R}^2 - k\lambda^2 T^a \eta_{ab} T^b +
2k^2 \lambda^4 + 2k\lambda^2 V^{\mu\nu} \tilde{R}_{\mu\nu}\right),\label{eq:covaction}
\ee
with
\be
\tilde{R}_{\mu\nu} := \partial_\mu \omega_\nu - \partial_\nu \omega_\mu
\ee
The last term in the action corresponds to the usual Einstein-Cartan term. In 2-dimensions
it is a total divergence and will be dropped. Note that when $k=0$, the above action
reduces simply to a single term, namely the curvature-squared term.

 The action (\ref{eq:covaction}) corresponds is of the same form as the Katanaev-Volovich model of 2-D gravity with torsion\cite{kv86,kv90s}, albeit with a specific ratio of coefficients determined by the gauge coupling parameter.
  
The equations of motion are the critical points of the action functional (\ref{eq:covaction}).
That is, since the the spin-connection $\omega_\mu$ and the frame-fields $e^a_\mu$ are functionally independent and
\be
\delta S=\frac{1}{\lambda^2}\int d^2x\sqrt{-g}\left(W^\mu \delta\omega_\mu+E^{a\mu}\delta e_{a\mu}\right),
\ee
we have that a necessary condition for a critical point is that
\bea
W^\mu&:=&-\nabla_\nu\tilde{R}^{\mu\nu}-C\epsilon_{ab} e^a_\nu T^{b\mu\nu}=0;\label{eq:omegaeom}\\
E^{a\mu}&:=&-C D_\nu T^{a\mu\nu}+e^a_\nu \tau^{\mu\nu}+\frac{C^2}{2}e^{a\mu}=0.\label{eq:eeom}
\eea
where we have defined $C:= k\lambda^2$. As well, the  above spacetime tensor indices $\mu,\nu,...$ are
raised and lowered by the `background metric' $g_{\mu\nu}:=\eta_{ab}e^a_\mu e^b_\nu$.  There are two
covariant derivatives.  The first, $\nabla_\nu$, is with respect to the background Lorentzian
metric $g_{\mu\nu}$, while the second, $D_\nu$ is with respect to the spin-connection.  That is
\be
D_\nu T^{a\mu\nu}:=\nabla_\nu T^{a\mu\nu}-\epsilon^a{}_b\omega_\nu T^{b\mu\nu}.
\ee
Finally, the tensor $\tau^{\mu\nu}$ is defined as
\bea
\tau^{\mu\nu}&:=&\tilde{R}_\pi{}^\mu \tilde{R}^{\pi\nu}+C\eta_{ab}T^a{}_\pi{}^\mu T^{b\pi\nu}\none
&-&\frac{1}{4}g^{\mu\nu}\left(\tilde{R}^2+C T^2\right),
\eea
where ${\tilde R}^2:=\tilde{R}_{\mu\nu}\tilde{R}^{\mu\nu}$ and $T^2:=\eta_{ab} T^a{}_{\mu\nu}T^{b\mu\nu}$.

\section{Hamiltonian Analysis}

We parametrize the `background metric' $g_{\mu\nu}=\eta_{ab}e^a_\mu e^b_\nu$, where the frame-field components
are the $J_a$ components of the gauge potential $A_\mu=\lambda e^a_\mu J_a +\omega_\mu J$.  We write:
\be
e^0=n dt+pdx;\qquad e^1=qN^1 dt +qdx.
\ee
We note that the metric is:
\be
g_{\mu\nu}= \left(
\begin{array}{c c}
-(n^2-q^2(N^1)^2)&-np+q^2N^1\\
 -np+q^2N^1&q^2-p^2
\end{array}
\right)
\ee
\be
\sqrt{-g}=q N,
\ee
where $N:=n-N^1 p$. Note also that
\be
g^{\mu\nu}= \frac{1}{q^2N^2}\left(
\begin{array}{c c}
-(q^2-p^2)&-np+q^2N^1\\
 -np+q^2N^1&(n^2-q^2(N^1)^2)
\end{array}
\right)
\ee
Another potentially useful form of the metric is:
\be
ds^2= -\frac{q^2N^2}{q^2-p^2}dt^2+ (q^2-p^2)\left(dx + \frac{q^2N^1- np}{q^2-p^2}dt\right)^2
\ee

As before, we define:
\be
F=\partial_0\omega_1-\partial_1\omega_0\, ,
\ee
 and
\bea
T^0 &=&\dot{p}-n'+q(\omega_1 N^1-\omega_0);\none
T^1 &=&\dot{q}-(qN^1)'-\omega_0 p +\omega_1 n.
\eea
The action is
\be
S_{YM}=\frac{1}{2\lambda^2}\int d^2x\left[\frac{1}{Nq}(F^2+ k\lambda^2 (T^1)^2-k\lambda^2 (T^0)^2)+\lambda^4N q\right].
\ee
Note that for the group ISO(1,1) (i.e. $k=0$) only the $F^2$ term remains. In two dimensions this
gives a rather trivial solution space so we henceforth consider only $k=\pm1$.\\

The momenta canonically conjugate to $p,q,n,N^1,\omega_0,\omega_1$ are respectively
\bea
\Pi_p&=&-\frac{k T^0}{ N q};\label{eq:conp}\\
\Pi_q&=&\frac{k T^1}{ N q};\label{eq:conq}\\
\Pi_n&=&0;\\
\Pi_1&=&0;\\
P_0&=&0;\\
P_1&=&\frac{F}{\lambda^2 N q}.
\label{eq:P1 F}
\eea

The total Hamiltonian density is `pure constraint':
\be
H=N H_s +N^1 D +\omega_0 M,
\ee
where the Hamiltonian constraint $H_s$ is
\be
H_s:=\frac{q}{2 }(-k\Pi_p^2+k\Pi_q^2+ \lambda^2 P_1^2) -\frac{1}{4} q -D\Pi_p,
\ee
the diffeo constraint $D$ is
\be
D:=-q D\Pi_q-p D\Pi_p,
\ee
where $D\Pi_q:=\Pi_q'+\omega_1 \Pi_p$ and $D\Pi_p:=\Pi_p'+\omega_1 \Pi_q$.  Finally the `Gauss law constraint' is
\be
M:=-P_1'+q\Pi_p+p\Pi_q.
\ee
Note that above, $\Pi_q':=\partial_x \Pi_q$ is the spatial derivative.

The self-consistency of these constraints, that is $0\approx\dot H_s(x)=[H_s(x),\int dy H(y)]$, etc., must be checked.
We smear the constraints: $H_s[u]:=\int dy u(y) H_s(y)$, etc., and find
\be
[H_s[u], H_s[v]]=0;
\ee
\be
[H_s[u],M[v]]=\int dx \frac{uv}{q}\left(p H_s - D\right)\approx 0;
\ee
\bea
[H_s[u], D[v]]&=&\int dx\left\{\lambda^2 u vq P_1 M +\frac{ u v\omega_1}{q} D -
\frac{v}{q}\left(qu'+u\omega_1 p\right)H_s\right\}\none
&\approx&0;
\eea
\be
[D[u], M[v]]=0;
\ee
and this is a strong equality.
\be
[D[u], D[v]]=D[u v'-v u']\approx0.
\ee
And finally
\be
[M[u], M[v]]=0,
\ee
strongly.

\bigskip\noindent
We see that the constraint algebra closes, and the constraints are self-consistent.

The equations of motion are:
\bea
\dot{\omega}_1 &=& \lambda^2 N q P_1 +\omega'_0;\\
\dot{q} &=& k N q \Pi_q-N\omega_1 +(qN^1)'-N_1p\omega_1+\omega_0 p;\\
\dot{p}&=&-Nk q \Pi_p + N' +(N^1p)'-N^1q\omega_1+\omega_0q;\\
\dot{P}_1 &=& N\Pi_q+N^1\left(q\Pi_p +p\Pi_q\right)\label{eq:P1dot}\\
\dot{\Pi}_q&=& -\frac{k}{2} N \Omega+ N^1 D\Pi_q-\omega_0 \Pi_p;\label{eq:Piqdot}\\
\dot{\Pi}_p &=& N^1 D\Pi_p - \omega_0 \Pi_q.\label{eq:Pipdot}
\eea
In the above we have defined:
\be
\tOmega := \frac{1}{2}\left(-k\Pi^2_p+k\Pi^2_q+ \lambda^2 P_1^2 -\lambda^2\right)
\ee

\section{Solutions}

The gauge is fixed by
\be
p\approx0,\omega_1\approx0,Q:=q-1\approx0.
\ee
The consistency conditions for this choice are, respectively:
\bea
&N'+\omega_0-{k} N \Pi_p\approx0;\label{eq:pdot}\\
&\omega_0'+{k} CN P_1\approx0;\label{eq:omega1dot}\\
&(N^1)'+{k} N \Pi_q\approx0.\label{eq:qdot}
\eea

The constraints reduce to
\bea
H_s&=&\frac{{k}}{2}\tOmega-\Pi_p'\approx0;\label{eq:Hsgf}\\
D&=&-\Pi_q'\approx0;\label{eq:Dgf}\\
M&=&\Pi_p-P_1'\approx0.\label{eq:Mgf}
\eea

Now from Eq.(\ref{eq:Dgf}) we have that $\Pi_q=\Pi_q(t)$ is an integration (spatial) constant.  We use
this and Eq.(\ref{eq:Mgf})(which allows us to replace $\Pi_p$ by $P_1'$) in Eq.(\ref{eq:Hsgf})
to get the second order differential equation
\be
P_1''+\frac{{k}}{2}(P_1')^2-\frac{\lambda^2}{2}P_1^2-B=0,
\label{eq:P1de}
\ee
where
\be
B:=\frac{{k}}{2}\Pi_q^2-\frac{\lambda^2}{2}.
\ee

There are two classes of solutions to (\ref{eq:P1de}). This can be seen as follows. Define:
\be
C_1:= e^{{k} P_1}\left[(P_1^\prime)^2 -\lambda^4(k P_1 -1)^2 +\Pi_q^2 \right]
\label{eq:C1}
\ee
It is easy to verify that
\be
C_1^\prime = \frac{1}{2}e^{{k} P_1} P_1^\prime
     \left[P_1''+\frac{{k}}{2}(P_1')^2-\frac{\lambda^2}{2}P_1^2-B \right]
\ee
Thus the solutions bifurcate into two classes:
\bea
P_1^\prime =0 &\rightarrow& \frac{\lambda^2}{2}P_1^2-B\\
P_1^\prime \neq 0 &\rightarrow& C_1 = C_1(t)
\eea
As we will see, the first condition requires that the torsion be zero. It leads to a
solution-space of lower dimension. The second condition allows for non-zero torsion.

\subsection{Torsion-less Solutions}

This class of solutions of the Hamiltonian constraint has $P_1'=0$, and hence
\be
P_1^2+\frac{k\Pi_q^2}{\lambda^2}=1.
\label{eq:P1 TF}
\ee
In this case the Gauss Law constraint $M=0$ implies $\Pi_p=0$.  If we now use these in the
equation of motion (\ref{eq:Pipdot}) for $\dot{\Pi}_p$, we find that either $\omega_0=0$ or $\Pi_q=0$.
If we use the former in the consistency condition  $0=\dot{\omega}_1=\omega_0'+\lambda^2 N P_1$, then
either $N=0$, which leads to a degenerate geometry, or $P_1=0$.  But $P_1=0$ implies $\dot{P}_1=0$,
and hence the equation of motion for $\dot{P}_1$ implies $\Pi_q=0$.  Hence we must have
that both $\Pi_p$ and $\Pi_q$ are zero;  that is, the metric is torsion-free.
Since $\Pi_q=0$ we now find from (\ref{eq:P1 TF}) above that $P_1^2=k^2=1$.

We now find that from the consistency conditions (\ref{eq:pdot}) and (\ref{eq:omega1dot}) that
\bea
&N'+\omega_0\approx0;\label{eq:pdot2}\\
&\omega_0'+k\lambda^2 N\approx0;\label{eq:omega1dot2}
\eea

 that $N''=k\lambda^2 N$ and $N^1=N^1(t)$.  The function $N$ is then of the form $N_0(t)\sin{{\lambda}(x-x_0(t))}$,
 respectively $N_0(t)\sinh{{\lambda}(x-x_0(t))}$, as $k>0$, respectively $k<0$.
 In these expressions, $N_0(t),x_0(t)$ are integration constants.  The metric is then of the form (with $k<0$):
\be
ds^2=-\left(N_0(t)\sinh{{\lambda}(x-x_0(t))}\right)^2dt^2+(dx+N^1(t)dt)^2.
\ee

We have not completely fixed the coordinate invariance. One can choose:
\be
dy = dx + N^1(t)dt
\ee

As well, the lapse can be set to one using the residual time reparameterization invariance
so that the metric becomes:
\be
ds^2=-\sinh^2{\left({\lambda}(y-y_0(t))\right)}dt^2+dy^2.
\ee
where
\be
y_0(t) := x_0(t) + \int dt N_1(t)
\ee

The remaining free function $y_0(t)$ is related to the fact that we have not completely fixed
the gauge invariance.  Indeed, the nontrivial consistency conditions, (60, 61),
for the torsionless case, where $\Pi_p=0,P_1=\pm 1/\alpha$ boil down to
(\ref{eq:pdot2}) and (\ref{eq:omega1dot2}), respectively, which can be written as a matrix equation
\be
\phi'=A\phi
\ee
where $\phi=[N,\omega_0]^T$ and
\[
A: = \left( \begin{array}{cc}
         0 & -1\\
        -k\lambda^2 & 0
\end{array}\right)
\]

The system is preserved under linear transformations $\phi\to \bar{\phi}=L\phi$, where $L$ is a 2x2 matrix:
\[
L = \left( \begin{array}{cc}
         b_1& b_2\\
        k\lambda^2 b_2& b_1
   \end{array}\right)
\]
which has unit determinant if $b_1^2-k\lambda^2 b_2^2=1$.  Note that $b_i$ are functions of $t$.
This is an $O(1,1)$ transformation.  We can use such a transformation to transform $\tilde{x}_0(t)$
away.  Indeed, such a transformation is given by
$b_1(t)=-b_0(t)\lambda\cosh{(\lambda y_0(t))},b_2(t)=b_0(t)\sinh{(\lambda y_0(t))}$.
This gives $N\propto \sinh{(\lambda y)}$.  Note that in this case $b_1^2-k\lambda^2 b_2^2=k\lambda^2 b_0^2(t)$.
On the other hand, if we choose to transform so that $N\propto \cosh{(\lambda y)}$,
then we would have different $b_1,b_2$ satisfying $b_1^2-k\lambda^2 b_2^2=-k\lambda^2 b_0^2(t)$.
Thus only one of these transformations is continuously connected to the identity transformation.

Now consider the stationary metric
\be
ds^2=-n^2(z)d\tau^2+q^2(z) dz^2.
\ee
For this to have constant curvature, that is, for $R=-2 \lambda^2$, it is required that
\be
q n''-q' n'- k\lambda^2 q^3 n=0.
\ee
If you solve this for $n(z)$ with $k=-1$ you get
\be
n(z)=A\cosh(\lambda (\theta))+B \sinh(\lambda (\theta)),
\ee
where in general the integration constants $A,B$ can be $\tau$-dependent.
In the above, $\theta'=q(z)$.  Now choose $y$ as a new coordinate, so that $\int dz q(z)=y - y_0(\tau)$.  The special case $A=0$ is then
\be
ds^2=-\sinh{(\lambda (y-y_0(t)))} dt^2 +dy^2.
\ee
where we have scaled $B$ away by a trivial coordinate transformation $dt=B(\tau)d\tau$.

The Ricci scalar for this metric is $R=-2\lambda^2$, a metric with constant negative curvature.
For $k>0$ we get the above, but with $\sinh$ replaced by $\sin$.  In this case
$R=2\lambda^2$, giving us a space of constant positive curvature.

This solution, as we have seen is torsionless, and of constant curvature.
In fact, the solution has a flat Yang-Mills connection, that is, $F^i_{\mu\nu}=0$.
Furthermore, the expression $\tau^{\mu\nu}$, which is quadratic in the Yang-Mills
curvature, satisfies $\tau^{\mu\nu}+\frac{\lambda^4}{2} g^{\mu\nu}=0$.

\subsection{ Torsion-full Solution}
One can verify that in the case $P_1^\prime \neq0$, the following is the first integral of Eq.(\ref{eq:P1de})
\bea
(P_1^\prime)^2 &=& \left[C_1 e^{-{k} P_1}+k\lambda^2({k} P_1-1)^2+\Pi_q^2\right]\nonumber\\
  &=:&f^2(t,P_1)
\label{eq:solx}
\eea
where $ C_1(t)$ is an integration constant.  We note also that as we have seen above, $\Pi_q'=0$ and hence $\Pi_q=\Pi_q(t)$.

Now it is easy see that $\Pi_p$ is given as a function of $P_1$ from Eq.(\ref{eq:Mgf}) by
\be
\Pi_p=P_1'=f(t,P_1)=\pm\frac{1}{{k}}\left[C_1 e^{-{k} P_1}+k\lambda^2({k} P_1-1)^2+ \Pi_q^2\right]^{\half}.
\ee

Differentiate (\ref{eq:pdot}) with respect to $x$.  We write, for notational ease $r:=P_1$.  We get
\be
N''+\omega_0'-{k}(Nf)'=0.
\ee
We replace $\omega_0'$ according to (\ref{eq:omega1dot}) and write $f'=\partial_x r f_r$ where $f_r=\partial_r f$ to get
\be
f^2N_{rr}+(f f_r-{k} f^2)N_r-{k}(f f_r+Cr)N=0.
\ee
The general solution is (according to MAPLE)
\be
N(t,r)=f e^{{k} r}(B_1(t) g+B_2(t)),
\ee
where $B_1(t),B_2(t)$ are integration constants and $g(t,r)$ is given by $g_r=e^{-{k} r}f^{-3}$.
We can now compute $\omega_0(t,r)$ from (\ref{eq:pdot}) to get
\be
\omega_0(t,r)= -f_r N-\frac{B_1(t)}{f}.
\ee

From (\ref{eq:qdot}) we find
\be
(N^1)_r=-{k}\Pi_q e^{{k} r} (B_1(t) g+B_2(t)).
\ee

Now consider the equations of motion for the time derivatives of the momenta:
\bea
\dot{r}&=&N \Pi_q+N^1 f; \label{eq:rdot}\\
\dot{\Pi}_q&=&-f(N  f_r+ \omega_0); \label{eq:piqdot}\\
\dot{f}&=&N^1 f f_r-\omega_0 \Pi_q.\label{eq:pipdot}
\eea

Consider the linear combination
\be
f_r\dot{r}-\dot{f}=(N f_r+\omega_0)\Pi_q,
\ee
by (\ref{eq:rdot}) and (\ref{eq:pipdot}).  Use (\ref{eq:piqdot}) on the right hand side to get
\be
\Pi_q\dot{\Pi}_q+f(f_r\dot{r}-\dot{f})=0\label{eq:sys}.
\ee

According to (\ref{eq:solx})
\be
\frac{d}{dt}f^2=\dot{r}(f^2)_r+\dot{C}_1(t) e^{-{k} r}+2\Pi_q\dot{\Pi}_q.
\ee
Thus using this in (\ref{eq:sys}) we get $\dot{C}_1=0$ and hence $C_1$ is both a space and time constant.

From (\ref{eq:piqdot}) we then get $\dot{\Pi}_q=B_1(t)$.

Consider again (\ref{eq:pipdot}).  After multiplying by $f$.  The left hand side becomes, after  using (\ref{eq:solx}) \be
\frac{d f^2}{dt}=2{k}\Pi_q\dot{\Pi}_q.
\ee
Hence (\ref{eq:pipdot}) becomes
\be
2{k}\Pi_q\dot{\Pi}_q=(f N^1+N\Pi_q)(f^2)_r +2B_1\Pi_q,
\ee
so that after canceling the left side with the last term on the right, we get using
the expressions obtained above for $N,\omega_0,N^1$:
\bea
0&=&f(f^2)_r\Pi_q\left[-{k} B_1 \int^r  du e^{{k} u} f(u)-B_2 e^{{k} r} +e^{{k} r}(B_1 g(r)+B_2)\right]\none
&=&f(f^2)_r\Pi_q\left[-{k} B_1 \int^r  du e^{{k} u} f(u) +e^{{k} r}B_1 g(r)\right]\none
&=& f(f^2)_r\Pi_q\int^r du f^{-3}(u).
\eea
where we integrated by parts to get the last equality.  Since we have already seen
that $B_1=\dot{\Pi}_q$, the most general nontrivial solution is $B_1=0$, so
that $\Pi_q$ is a spatial and temporal constant.

To compute the shift vector $N^1$, we solve (\ref{eq:qdot}).  The solution
contains an arbitrary function of time, but this in turn is
required to be zero by (\ref{eq:rdot}).

We now change the spatial coordinate from $x$ to $r$, so that $\dot{r}=0$.
We also change to an new time coordinate $\tau$ by $d\tau=B_2(t) dt$.
Then all the equations of motion and constraints are satisfied by
\bea
 f^2&=&C_1 e^{-{k} r}+k\lambda^2({k} r-1)^2+ \Pi_q^2;\\
N(\tau,r)&=&f e^{{k} r};\\
\omega_0(\tau,r)&=&-f f_re^{{k} r};\\
N^1(\tau,r)&=&-\Pi_q e^{{k} r}.
\eea

The metric is
\be
ds^2=-N^2d\tau^2+\frac{1}{f^2}\left(dr+f N^1 d\tau\right)^2.
\ee

The Ricci scalar and torsion of the above are, respectively:
\bea
R&=&-2k\lambda^2( r^2+{k} r -1)\\
T^0&=&-kNq\Pi_p= -ke^{kr}\left( C_1 e^{-{k} r}+k\lambda^2({k} r-1)^2+ \Pi_q^2\right) \\
T^1&=&Nq \Pi_q= \Pi_q e^{kr}\sqrt{C_1 e^{-{k} r}+k\lambda^2({k} r-1)^2+ \Pi_q^2}
\eea

Thus the general solution with torsion depends on two integration
constants, $C_1$ and $\Pi_q$. Event horizons exist for negative $C_1$.



\section{Conclusions}

We have presented a study of DS/ADS/Poincare Yang-Mills gravity. 
As in general relativity, the theory is background independent, although this is
done at the expense of reducing the symmetry group.  We have shown that
test `Higgs particles' traverse geodesics with respect to the Lorentzian
geometry determined by $e^i_\mu$.

In two spacetime dimensions the action is a special case of the Katanaev-Volovich model. We completed the Hamiltonian analysis
of the vacuum theory, confirming the existence of a generalized Birkhoff
theorem: the solutions are static and parametrized by two parameters.

In addition one of us (JG) is working on Yang-Mills gravity in 4D with
gauge group SO(4,2), in collaboration with S. Rahmati and S. Seahra.
In most work along these lines (see e.g. \cite{guo,wheeler}), the
torsion is forced to be zero {\it ab initio}.  We will relax this,
and explore implications, especially for cosmology.

We close by mentioning that there is another possibility in principle allows the
construction of an action that is invariant under the full gauge group.  One can
introduce two metrics:  one is dynamical, and determined by the
gauging $e^a_\mu$ of the generator $J_a$.  The other metric, the
background $g_{\mu\nu}$ is chosen in a way informed by
the uniformization theorems in 2 and 3D.  That is, given the
topology, the manifold will admit a particular `round' or homogeneous metric.
The idea is to choose the background to be precisely that round geometry.
This procedure is well defined in two or in three dimensions, but there is a problem in four or
higher dimensions, for which there is no known uniformization theorem.   Recall
that the 3D uniformization theorem was proved  using the Ricci flow \cite{thurston}.  The latter exists
in any dimension, and always converges to its fixed points,  the homogeneous
geometries.   So one could require that the  background geometry is
such that the Ricci flow of the geometry determined by the frame fields
and spin connection converges to it in the infinite limit of the flow
parameter.  However, given that we are really interested in working
with YM type actions, it is more sensible to postulate that the
consistency is provided by requiring that the Yang-Mills flow of
the gauge potential $A$ determined by the frame fields and spin
connection flows to the background `round geometry'.  In future
work we will therefore consider an alternate theory wherein the
background a metric which is the appropriate homogeneous geometry for some topology.

\section*{Acknowledgments }

The authors are grateful to Daniel Grumiller for drawing our attention to the Katanaev-Volovich model. We thank Sanjeev Seahra for useful conversations.
One of, JG, acknowledges the support of the CECS in Valdivia, Chile, where part of this work was done.  We also acknowledge the partial financial support of NSERC as well as  the Perimeter
Institute for Theoretical Physics (funded by Industry Canada and the Province
of Ontario Ministry of Research and Innovation).\\

\vspace{0.5in}



\begin{thebibliography}{99} 


\bibitem{townsend} P.K. Townsend {\it Small-scale structure of spacetime as the
origin of the gravitational constant}, Phys. Rev. {\bf D15}, 2795 (1977).

\bibitem{weyl} H.~Weyl, {\it Space-Time-Matter}, Methuen, London (1918).

\bibitem{early}  A sample, not exhaustive, list of papers on Yang-Mills gravity from the 70's and 80's:
 R.~ Utiyama, `Invariant theoretical interpretation of interaction',
 Physical Review {\bf 101},1597 (1956);doi:10.1103/PhysRev.101.1597;\\
 T. ~W. ~B. ~Kibble, `Lorentz invariance and the gravitationa
l field',
J. Math. Phys.
{\bf 2},212 (1961);
gravitational constant', Phys. Rev. D {\bf 15}, 2795 (1977).
S. ~W. ~MacDowell and F. ~Mansouri, `Unified geometric theor
y of gravity and su-
pergravity',
Phys. Rev. Lett.
{\bf 38}
, 739742 (1977);\\
K. ~Hayashi and T. ~Shirafuji, `Gravity from Poincar ´e gaug
e theory of fundamental
interactions',
Prog. Theor. Phys.
{\bf 64}
, 866882 (1980);\\
E.~A.~Ivanov and J.~Niederle, `Gauge formulation of gravitational theories. I.
The Poincare, de Sitter, and conformal cases', Phys. Rev.
D {\bf 25}, 976 (1981).

\bibitem{wheeler}  Again a sample of the literature in this area would include:
J.~T.~Wheeler, `Auxiliary field in conformal gauge theory', Phys. Rev. d {\bf 44}, 1769 (1991);
de Sitter gravity', Class. Quant. Grav. {\bf 24}, 4009 (2007).
or on `conformal gravity'.

\bibitem{guo} H.-Y ~Guo, et. al., `Snyder's model-de Sitter special relativity duality and
de Sitter gravity', Class. Quant. Grav. {\bf 24}, 4009 (2007); C.-G. Huang,
H.-Q. Zhang, H.-Y. Guo: {\it Cosmological solutions with torsion in a
model of the de Sitter gauge theory of gravity}.  JCAP 10 (2008) 010;
C.-G. Huang, M.-S. Ma: On torsion-free vacuum solutions of the model of de Sitter
gauge theory of gravity (II). Front. Phys. China, 4 (2009) 525–529;
C.-G. Huang, M.-S. Ma: {\it de Sitter spacetimes with torsion in the model of de Sitter
gauge theory of gravity}. Phys. Rev. D 80 (2009) 084033;
C.-G. Huang, M.-S. Ma: {\it A new solution with torsion in model of dS gauge theory of
gravity}. Commun. Theor. Phys. 55 (2011) 65–68.
%

\bibitem{ymflow} J.~ Gegenberg, A.~C. ~Day, H.~ Liu and
S.~ S. ~Seahra
`An instability of hyperbolic space under the Yang-Mills flow'
Journal of Mathematical Physics,
{\bf 55}
, 042501
arXiv: 1210.0839 [hep-th].
%

\bibitem{kv86}M. O. Katanaev and I. V. Volovich, “String model with dynamical geometry and torsion,” Phys. Lett. B175 (1986) 413–416. [arXiv:hep-th/0209014]. 

\bibitem{kv90s} 
P. Schaller and T. Strobl, ``Canonical Quantization of Non-Einsteinian Gravity and the Problem of Time'',  Class. Quant. Grav.11 (1994) 331-346 [hep-th/9211054]
 ;
Noriaki Ikeda and Ken-Iti Izawa, ``Quantum Gravity with Dynamical Torsion in Two Dimensions'' Prog. Theor. Phys. 89 (1993) 223-230;
T. Strobl, ``Comment on Gravity and  the Poincare Group'', 	Phys.Rev. D48 (1993) 5029-5031  [arXiv:hep-th/9302041];
  W. Kummer and D. J. Schwarz, “General analytic solution of R**2 gravity with dynamical
torsion in two-dimensions,” Phys. Rev. D45 (1992) 3628–3635.
   M. O. Katanaev, W. Kummer, and H. Liebl, “Geometric interpretation and classification of global solutions in generalized dilaton gravity,” Phys. Rev. D53 (1996) 5609–5618  [gr-qc/9511009].
%
\bibitem{jt} R. Jackiw, in
Quantum Theory of Gravity
, edited by S. Christensen
(Hilger, Bristol, 1984), p. 403;C. Teitelboim, in
Quantum Theory of
Gravity
, edited by S. Christensen (Hilger, Bristol, 1984), p.327; M
. Hen-
neaux, Phys. Rev. Lett.
54
, 959 (1985).  See also D. Louis-Martinez, J. Gegenberg and G. Kunstatter, Phys
. Letts.
B {\bf 321}
,
193 (1994).

%
\bibitem{wigner}   E. ~Inönü, E.P. ~Wigner,
{\it On the Contraction of Groups and Their Representations},
Proc. Nat. Acad. Sci. {\bf 39} (6): 51024 (1953).

\bibitem{spherym} O. ~Brodbeck and N.~Straumannm
{\it A generalized Birkhoff theorem for the Einstein-Yang-Mills system}, J. Math. Phys. {\bf 34}, 2412-2423 (1993);
T. A. ~Oliynyk and H. P. ~Kunzle, {\it On all possible static
spherically symmetric EYM solitons and black holes}, Class. Quant. Grav. {\bf 19}, 457 (2002), [arXiv:gr-qc/0109075].

\bibitem{thurston} The original geometrization conjecture of W.P. Thurston was proved by J. Hamilton and G. Pereleman.  For references, see J.~ W. ~Morgan. {\it Recent progress on the Poincaré conjecture and the classification of 3-manifolds}, Bulletin Amer. Math. Soc. 42 (2005) no. 1, 57-78.


\end{thebibliography}
\end{document}